\documentstyle[prl,aps,epsf,multicol,tabularx]{revtex}
\begin{document}
\draft
\title{A Model for the Voltage Steps in the Breakdown of the Integer Quantum Hall Effect}
\author{A. M. Martin, K.A. Benedict, F.W. Sheard and L. Eaves}
\address{School of Physics and Astronomy, University of Nottingham,
Nottingham, NG7 2RD, UK.}
\date{\today}
\maketitle
\begin{abstract}
In samples used to maintain the US resistance standard the
breakdown of the dissipationless integer quantum Hall effect
occurs as a series of dissipative voltage steps. A mechanism for
this type of breakdown is proposed, based on the generation of
magneto-excitons when the quantum Hall fluid flows past an ionised
impurity above a critical velocity. The calculated generation rate
 gives a voltage step height in good agreement
with measurements on both electron and hole gases. We also compare
this model to a hydrodynamic description of breakdown.
\end{abstract}
\begin{multicols}{2}
%
%
 In the integer quantum
Hall effect (IQHE) regime \cite{Klitzing}, a two-dimensional
electron fluid carries an almost dissipationless current and the
ratio of the current, $I_x$, to the Hall voltage, $V_H$, is
quantized in units of $e^2/h$. However, above a critical current,
the dissipative voltage, $V_{x}$, measured along the direction of
current flow, increases rapidly, leading to quantum Hall breakdown
(QHBD). Several possible mechanisms for QHBD \cite{Natchwei} have
been proposed: avalanche heating \cite{Komiyama1985}; percolation
due to an increase of delocalized states \cite{Trugman};
quasi-elastic inter-Landau level scattering
 \cite{Eaves1984,Heinonen,Eaves1986}; acoustic
phonon emission due to intra-Landau level scattering
\cite{Streda}; formation of compressible metallic filaments
\cite{Tsemekhman}; and resonant impurity scattering of electrons
 \cite{Pokrovsky}. For certain samples, including
those used to maintain the US resistance standard at the National
Institute of Standards and Technology (NIST), breakdown occurs as
a series of up to twenty steps, in $V_x$, of roughly equal height
$\Delta V_x \simeq 5mV$, see Fig. 1 \cite{Cage1983,Lavine}. This
is fundamentally different from the case in which breakdown is
observed as a single large increase of $V_x$ \cite{Komiyama1985}.

IQHE breakdown is not only of fundamental interest, but is also
relevant to quantum metrology since a large value of $I_x$ can
improve the measurement precision. In this work we develop a
theoretical model to account for the dissipative steps observed by
the NIST group \cite{Cage1983,Lavine} and others
\cite{Blieketal,Eaves2000}. We show that, in the presence of
charged impurity-induced disorder, the quantum Hall fluid (QHF) is
unstable when the local fluid velocity exceeds a critical value.
Under these conditions, magneto-exciton or electron-hole (e-h)
pair excitations
 are generated spontaneously near
an impurity. The voltage step height, $\Delta V_x$, is directly
related to the rate of formation of the pairs, which we calculate
using a parameter-free model. This type of excitation of the QHF
is analogous to vortex-antivortex pair formation in classical or
quantum fluid flow around an obstacle \cite{Eaves1999}.

Earlier work \cite{Igor,KandH,Mac} has formulated a method for
calculating the excitation dispersion relation of e-h pairs
generated by exciting an electron from an occupied Landau level
$n$ to an unoccupied Landau level $(n+1)$, where $n$ is the Landau
level index. By extending these models
\begin{figure}
\epsfxsize=6.5cm \centerline{\epsffile{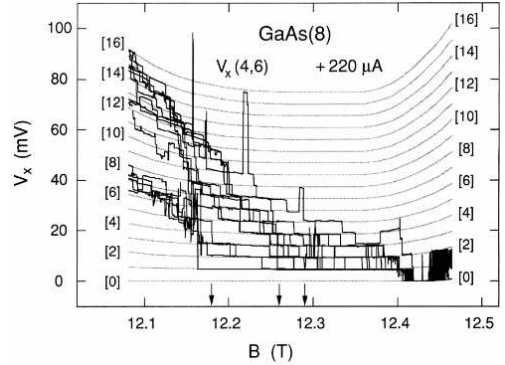}}
\vspace{0.0cm} \caption{Step like breakdown in the integer quantum
Hall effect, from Ref. [12]. Each step contributes a multiple of
$\simeq 5mV$ to the dissipative longitudinal voltage.}
\end{figure}
\noindent to include the electric field, arising from a charged
impurity and the Hall voltage drop across the sample, we obtain a
critical electric field at which it costs no energy to generate an
e-h pair at a given wave-vector $Q$. We incorporate this into a
calculation of the generation rate, $W$, of these pairs
\cite{Chaubet,Ando} due to a single charged impurity. This then
gives a dissipative voltage increment, $\Delta V_x=hW/(2e)$.

Our starting point is the Fermi golden rule, from which we
calculate the rate of generation of e-h pairs, in the Landau
gauge, due to a single charged impurity in a 2DEG
\cite{Chaubet,Ando},
\begin{eqnarray}
& &W_{n,(n+1)}= \frac{2 \pi}{\hbar}
\delta(\epsilon_{n}-\epsilon_{(n+1)}) \times \nonumber
\\ & &  \left| \int_{r} d^3r
\left|\Xi_0(z)\right|^2\phi_{n}(r_{\perp},k_x)\phi_{(n+1)}(r_{\perp},k_x^{\prime})V(r)\right|^2
,
\end{eqnarray}
where $V(r)$ is the impurity Coulomb potential,
$\phi_{n}(r_{\perp},k_x)$ is the electronic eigen-function in the
$x$$y$ plane and $\Xi_0(z)$ is the envelope function of the first
electronic sub-band. We assume that the $n$ lower Landau levels
are filled and the ($n+1$) level is empty. There is thus no static
screening of the impurity charge \cite{Tamura}. This gives
\begin{eqnarray}
& &W_{n,(n+1)}= \frac{2 \pi}{\hbar}
\delta(\epsilon_{n}-\epsilon_{(n+1)})\left| \int_{q_{\perp}} d^2
q_{\perp}\frac{S}{(2\pi)^2} \times \right.\nonumber
\\ & &  \left. \int_{r_{\perp}} d^2r_{\perp}
\phi_{n}(r_{\perp},k_x)\phi_{(n+1)}(r_{\perp},k_x^{\prime})e^{iq_{\perp}.r_{\perp}}
F_i(q_{\perp},z_i)\right|^2,
\end{eqnarray}
where
\begin{equation}
F_i(q_{\perp},z_i)=\frac{e^2}{2 S \kappa q_{\perp}}
\int_{-\infty}^{\infty}\left|\Xi_0(z)\right|^2
e^{-q_{\perp}\left|z-z_i\right|} dz.
\end{equation}
Here $\kappa$ is the dielectric constant (for GaAs $\kappa=0.11nF
m^{-1}$). If we assume that the impurity is at the center of the
subband then,
\begin{eqnarray}
& & W_{n,(n+1)}= \frac{2 \pi}{\hbar}\left(\frac{e^2 2 \pi}{8 \pi^2
\kappa L_x}\right)^2 \delta(\epsilon_{n}-\epsilon_{(n+1)})
\nonumber
\\& & \left| \int_{q_{y}} dq_{y}
\frac{1}{q_{\perp}(1+\frac{q_{\perp}}{b})^3}I_{{n},(n+1)}^{k_x,k_x^{\prime}}\right|^2,
\end{eqnarray}
where
\begin{equation}
I_{n,(n+1)}^{k_x,k_x^{\prime}}  = \int_y dy
\Phi_{n}(y-Y_{k_x})e^{iq_y y}\Phi_{(n+1)}(y-Y_{k_x^{\prime}}),
\end{equation}
$\Phi_{n}(y-Y_{k_x})$ is the simple harmonic oscillator solution
to the Schr\"{o}dinger equation centered on $Y_{k_x}=l_B^2k_x-E_y
m/(eB^2)$, $m$ is the mass of the electron, $E_y$ is the $y$
component of the electric field, $l_B=\sqrt{\hbar/eB}$ is the
magnetic length and $B$ is the magnetic field.
 Calculating the transition rate out of state ($n$,
$k_x$) we find that
\begin{eqnarray}
& & W_{n}= 2\sum_{k_x^{\prime}}W_{n,(n+1)}=\frac{2L_x}{2 \pi}
\int_{k_x^{\prime}}dk_x^{\prime}W_{n,(n+1)} = \nonumber
\\ &&
 \frac{1}{L_x \hbar} \left(\frac{e^2 }{4 \pi \kappa}\right)^2 \left| \frac{\partial k_x}{\partial \Delta
\epsilon} \right|_{\Delta \epsilon=0}  \left| \int_{q_{y}} dq_{y}
\frac{1}{q_{\perp}(1+\frac{q_{\perp}}{b})^3}
I_{n,(n+1)}^{k_x,k_x^{\prime}}\right|^2, \nonumber
\\ & &
\end{eqnarray}
where \cite{Igor,KandH,Mac}
\begin{equation}
\Delta \epsilon = \hbar \omega_c+\frac{e^2}{4 \pi \kappa l_B}
\Delta_{n,(n+1)}[k_x^{\prime}-k_x]-eE_yl_B^2(k_x^{\prime}-k_x)
\end{equation}
and $\Delta_{n,(n+1)}[k_x^{\prime}-k_x]$ includes the exchange and
Coulomb local-field corrections, which are independent of $E_y$.
To obtain Eq. (7) we have assumed $e^2/\kappa a_L \ll \hbar
\omega_c$; hence we neglect Landau level mixing. In Eq. (6) there
is an implicit condition for $k_x^{\prime}$ which must be
calculated. This condition can be obtained from Eq. (7) when
$\Delta \epsilon=0$. If exchange and Coulomb interactions are
omitted this condition is simply the same as for energy-conserving
elastic inter-Landau level transitions. However, when interaction
terms are included $Q=k_x^{\prime}-k_x$ must be evaluated
numerically. Eq. (7) can be split into two components: the
excitation interaction energy, given by the first two terms, which
is independent of $E_y$ and the electrostatic energy, $eE_yl_B^2
Q$. In Fig. 2 the crossing point of the electrostatic energy
(dashed line) and the excitation inter-
\begin{figure}
\epsfxsize=6.0cm \centerline{\epsffile{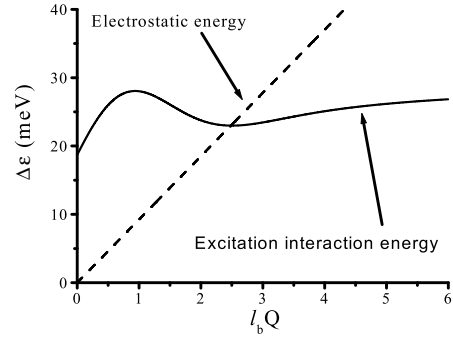}} \vspace{0.0cm}
\caption{The magneto-exciton mode energy, $\hbar \omega_c+e^2/(4
\pi \kappa l_B) \Delta_{0_{\sigma},(1)_{\sigma}}[Q]$ (solid line)
for excitations from $n=0$ to $n=1$, for $E_y=0$; the dashed line
is the electrostatic energy, $eE_yl_B^2Q$. $B=12.3T$ and $E_y=1.5
\times 10^{6}V/m$.}
\end{figure}
\vspace{-0.3cm} \noindent action energy (solid line) gives the
value of the e-h separation ($l_B^2Q$) for which it costs no
energy to generate e-h pairs, i.e. $\Delta \epsilon=0$. To obtain
the total rate of production of e-h pairs, we take Eq. (6) and sum
over all initial states such that
\begin{equation}
W = 2\sum_{k_x} W_{n}=2 \frac{L_x}{2 \pi} \int_{k_x}dk_x W_{n}.
\end{equation}
\vspace{-0.0cm}
 \noindent The above equation gives us the
generation rate of e-h pairs by a single charged impurity at given
$B$ and $E_y$.

Consider a local region of the sample where $E_y$ is large enough
to create e-h pairs, due to scattering from a charged impurity, at
a rate given by Eq. (8). Such regions can be expected to occur at
high current, possibly near the sample edge where the Hall field
is expected to be large \cite{Mac2}. A pair created close to an
impurity will drift along the Hall bar at a velocity $v \approx
E_y/B$, so one can imagine, for a fixed generation rate, a stream
of e-h pairs moving along the Hall bar. Then, for $\nu=2$, we have
a situation in which a small fraction of electrons in the lower
Landau level ($n=0$) have been replaced by holes and the
previously empty upper Landau level ($n=1$) contains some
electrons. As the e-h pairs move away from the high field region,
the spacing between the electron and hole in a pair will increase
and most pairs will eventually ionise by acoustic phonon emission.
Due to the absence of empty states into which the excited electron
can relax and neglecting weak, second order Auger processes, we
can assume that all the generated e-h pairs will eventually ionise
and lead to a dissipative current $i=e W$, flowing across the Hall
voltage equipotentials.  At $\nu=2$ this gives a dissipative
voltage
\begin{equation}
V_{x}= \left(\frac{hW}{2e}\right).
\end{equation}

We now compare the results of our model with breakdown
measurements which show voltage steps. The NIST experiments on the
US resistance standard samples \cite{Lavine} were carried out at
$\nu=2$ and $B=12.3T$. The experi-
\begin{figure}
\epsfxsize=8.5cm \centerline{\epsffile{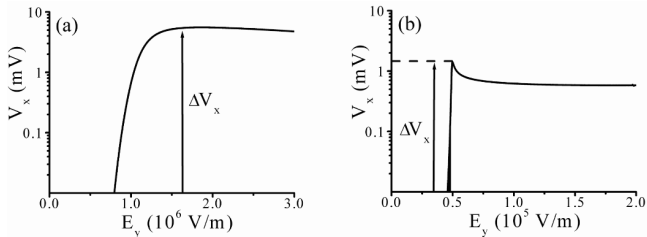}}
\vspace{-0.0cm} \caption{The dissipative voltage, for electron
[12] (a) and hole [14] (b) gas samples, calculated from Eq. (9) as
a function of $E_y$. The parameters used refer to the experimental
conditions of [12,14]: (a) $m^{\star}=0.07m_e$, $N_s=3
\times10^{15}m^{-2}$ and $B=12.3T$; (b) $m^{\star}=0.4m_e$,
$N_s=10^{15}m^{-2}$ and $B=2.1T$. }
\end{figure}
\vspace{-0.3cm}  \noindent ments show a series of dissipative
steps in $V_x$ of regular height $\Delta V_x \simeq 5mV$, see Fig.
1. In Fig. 3(a) we plot our calculated dissipative voltage as a
function of the background electric field $E_y$.

Since the rate of production of e-h pairs is strongly influenced
by the overlap between the wave-functions in the occupied ($n=0$)
and unoccupied ($n=1$) Landau levels, it increases rapidly at a
{\it critical} electric field. This critical electric field occurs
when Eq. (9) gives a $V_x$ comparable to the small background
dissipative voltage governed by $\sigma_{xx}$. The rate, and hence
$V_x$, then reaches a maximum when the electric field is such that
the e-h pairs are formed close to the roton minimum of the
magneto-exciton dispersion curve, $Q l_B \approx 2.5$. For a given
$E_y$, for which the rate of production of e-h pairs is finite,
the pairs formed at the breakdown point will relax, i.e. electrons
in the upper Landau level relax their energy by moving towards one
side of the Hall bar, whilst holes move in the opposite direction.
This process tends to screen the Hall field over much of the Hall
bar. Since the Hall voltage in the NIST experiments remains
constant at its quantized value over the magnetic field range over
which the dissipative steps occur, this screening effect tends to
enhance the electric field at the breakdown point. Thus, as the
{\it critical} electric field is reached, the generation rate at
the breakdown point increases rapidly, inducing a further increase
in $E_y$, due to the screening of the Hall field in other regions
of the Hall bar. For breakdown at a single charged impurity we
therefore expect the system to switch between two stable states,
corresponding to $V_x=0$ and $V_x \approx 5.6mV$, which
corresponds to the maximum value of $V_x$ in Fig. 3(a). In the
NIST experiments, a series of steps is observed, as seen in Fig. 1
and we attribute each step to the formation of separate streams of
e-h pairs generated by other charged impurities, i.e. each step is
associated with the local electric field at a particular impurity
reaching its {\it critical} value. Fig. 3(b) shows the results of
a similar calculation for the breakdown of the QHE observed in a
hole gas sample \cite{Eaves2000}. In these experiments the step
height was found to be $\Delta V_x \approx 1mV$. Our calculation
for this case gives a step height $\Delta V_x=1.6mV$, in
qualitative agreement with experiment.

The above calculated values are derived from a model which
combines a calculation of the magneto-exciton dispersion with an
impurity-related tunneling rate. We have included explicitly the
mechanism for the tunneling between the Landau levels and exchange
and Coulomb local field corrections.

An earlier paper by one of us \cite{Eaves1999} drew an analogy
between the process described here and the formation of vortices
behind an obstacle moving relative to a fluid (e.g. the von Karman
vortex street in classical hydrodynamics). Using our model we now
examine this analogy more closely. One of the key ideas in
understanding the quantum Hall states was the appreciation by
Laughlin \cite{Laughlin} that they have low Coulomb energy because
the particles are tied to zeros in the many-body wavefuntion.
Zhang, Hansson and Kivelson \cite{Z1} effected such a binding of
zeros to particles using a Chern-Simons construction. This leads
to an effective field theory for the fractional quantum Hall
states in which the quanta of flux of a fictitious magnetic field
are tied to each particle. This has two effects: firstly, in the
mean field approximation the fictitious flux cancels the real flux
through the system; secondly, the charge-flux composites become
bosonic. Hence, at the mean field level, the QHF in a field
corresponding to integer filling is replaced by a system of
composite bosons in zero magnetic field. Such a system forms a
charged superfluid and one can view the quantum Hall state as a
composite Bose condensate. From this, Stone \cite{Stone}
formulated an effective superfluid hydrodynamic model. The charged
elementary excitations, e-h pairs, of the QHF appear naturally in
this description as vortices in the order parameter for the
composite boson superfluid. In this language, our model for the
QHBD is the spontaneous creation of vortex-antivortex pairs when
the QHF fluid velocity around an impurity reaches a critical value
\cite{WenZee}.

According to Stone \cite{Stone} the equation of motion of the QHF
is \vspace{-0.2cm}
\begin{equation}
m^{\star}\left[{\bf \dot{v}}-\left[{\bf v} \times {\bf
\Omega}\right]\right]=e\left({\bf E}+{\bf v} \times {\bf
B}\right)-{\bf \bigtriangledown}
\left(\frac{m^{\star}}{2}\left|{\bf v}\right|^2+\mu\right),
\end{equation}
where ${\bf v}$ is the velocity field, $\mu$ is the local chemical
potential containing all the interaction terms and ${\bf \Omega}$
is the fluid vorticity given by
\begin{equation}
\Omega_x=\Omega_y=0  \, \,  {\rm and} \, \,
\Omega_z=\frac{\partial v_y}{\partial x}-\frac{\partial
v_x}{\partial y}.
\end{equation}

Using Eq. (11), in conjunction with the continuity equation for
the density of the QHF, and examining small pertubations in the
velocity field, of the form
\begin{equation}
v_x=\frac{E_y}{B}+\epsilon_1 {\rm cos}(Q x-\omega t)
\end{equation}
\vspace{-0.3cm} \noindent and
\begin{equation}
v_y=\epsilon_2 {\rm sin}(Q x-\omega t)
\end{equation}
\noindent we find that when ${\bf \bigtriangledown} \mu =0$,
\begin{equation}
\hbar \omega = \Delta \epsilon=-eE_yl_B(l_BQ)+\hbar \omega_c.
\end{equation}
This result is equivalent to Eq. (7) in the absence of
interactions and for $\Delta \epsilon=0$ corresponds exactly to
the elastic inter-Landau level tunneling condition
\cite{Eaves1986}. Alternatively within this hydrodynamic model it
corresponds to the condition required to generate a
vortex-antivortex pairs at zero energy.

To make a direct comparison with our earlier quantum mechanical
calculation we need to evaluate the dissipative voltage drop along
the sample due to the generation of these vortex-antivortex pairs
from a single impurity. For a specific system we would have to
rely on a numerical simulation of Eq. (10). However, we can make
considerable progress by implementing what we already know about
fluid mechanics \cite{Massey}. Consider an obstacle in the path of
a fluid. At a low fluid velocity the flow around the obstacle is
laminar. When the flow rate is increased vortex-antivortex pairs
are formed in the vicinity of the obstacle. However, a vortex
street is not formed until the flow is fast enough to free the
vortex-antivortex pairs from the local flow field near the
obstacle. In this steady state each vortex-antivortex pair moves
away from the obstacle at a velocity which is governed by the
background fluid velocity. This analogy suggests that the
vortex-antivortex pair in a QHF  moves away from the impurity at a
velocity given by $E_y/B$. From classical hydrodynamics
\cite{Massey} it is also known that the distance between each
vortex-antivortex ($l$) pair generated is approximately three
times the separation between a single vortex and antivortex ($d$)
($d/l=0.28$). Now consider two states for our fluid: firstly, the
state where the fluid flow is laminar; then we know that the
generation rate of vortex-antivortex pairs is zero. For the state
where we have a street of vortex-antivortex pairs we have the
condition $\Delta \epsilon=0$, hence from Eq. (14) we can evaluate
$E_y/B$. Dividing this velocity by the distance between each
vortex-antivortex pair ($d=0.28 l_B(l_BQ)$), to obtain the rate of
generation of vortex-antivortex pairs, we can use this
hydrodynamical model to estimate the voltage drop along the Hall
bar as
\begin{equation}
V_x=\frac{0.28 \pi \hbar \omega_c}{ e (l_BQ)^2}.
\end{equation}
To estimate the dissipative voltage step height from this
hydrodynamic model we have to evaluate Eq. (15). At the moment we
can only do this by referring back to the previous quantum
mechanical calculation to give us an estimate of $Ql_B$. Taking
the value for $Ql_B$ for which the quantum calculations give the
maximum value for $V_x$ we find, using Eq. (15) that $\Delta
V_x=4.8mV$ for the electron gas ($Ql_B \approx 1.9$) and $\Delta
V_x=0.75mV$ for the hole gas ($Ql_B \approx 1.1$).

Above we have shown both a quantum calculation of the dissipative
voltage drop along a Hall bar due to the presence of a charged
impurity and an effective fluid model for QHF which also predicts
the voltage drop along a Hall bar due to an impurity when a
critical fluid velocity is reached. Both of these calculations
predict voltage steps which agree with experiment. We believe that
the fluid model can be improved upon through numerical modeling
and the inclusion of interactions, however, the simple analysis
provided above gives a strong clue that future work in this
direction could be very useful in making direct comparisons with
experiments.

This work was supported by the EPSRC. We thank  V.A. Volkov, O.
Makarovsky,  A.C. Neumann, P.C. Main, and  T.M. Fromhold  for
useful discussions. We are grateful to NIST for use of the data
shown in Fig. 1.

\end{multicols}
\end{document}